\documentclass[a4paper]{jpconf_w_numbers}
\usepackage{graphicx}
\usepackage{hyperref}
\usepackage{amssymb}
\usepackage{amsmath}
\usepackage{xspace}
\hypersetup{colorlinks,urlcolor=blue,linkcolor=blue,citecolor=blue,filecolor=blue}
\newcommand{\NORSE}{\texttt{NORSE}\xspace}
\newcommand{\CODE}{\texttt{CODE}\xspace}
\newcommand{\LUKE}{\texttt{LUKE}\xspace}
\newcommand{\HS}{heat sink\xspace}
\newcommand{\sub}[1]{_{\mathrm{#1}}}
\renewcommand{\d}{\mbox{d}}
\newcommand{\dthreep}{\d^3 p}

\newcommand{\ED}{\ensuremath{E\sub{D}}}
\newcommand{\EDeff}{\ensuremath{E\sub{D,eff}}}
\newcommand{\EED}{\ensuremath{E/\ED}}
\newcommand{\EEDeff}{\ensuremath{E/\EDeff}}
\newcommand{\ESA}{\ensuremath{E\sub{SA}}}

\newcommand{\zeff}{Z\sub{eff}}
\newcommand{\epso}{\epsilon_{0}}
\newcommand{\lnL}{\ln\Lambda}
\newcommand{\me}{m\sub{e}}
\newcommand{\pder}[2]{\ensuremath{\frac{\partial #1}{\partial #2}}}

\newcommand{\power}[1]{\!\cdot\! 10^{#1}}

\newcommand{\Fig}[1]{Fig.~\ref{#1}}
\newcommand{\Ref}[1]{Ref.~\cite{#1}}

\newcommand{\changed}[1]{#1} 

\begin{document}
\title{Runaway-electron formation and electron slide-away in an ITER post-disruption scenario}

\author{A Stahl$^1$, O Embr\'eus$^{1}$, M Landreman$^{2}$, G Papp$^{3}$ and T F\"ul\"op$^{1}$}

\address{$^1$ Department of Physics, Chalmers University of Technology, G\"oteborg, Sweden}
\address{$^2$ Institute for Research in Electronics and Applied Physics, University of Maryland, College Park, Maryland 20742, USA}
\address{$^3$ Max Planck Institute for Plasma Physics, Garching, Germany}

\ead{stahla@chalmers.se}


\begin{abstract}
Mitigation of runaway electrons is one of the outstanding issues for the reliable operation of ITER and other large tokamaks, and accurate estimates for the expected runaway-electron energies and current are needed. Previously, linearized tools (which assume the runaway population to be small) have been used to study the runaway dynamics, but these tools are not valid in the cases of most interest, i.e.~when the runaway population becomes substantial. We study runaway-electron formation in a post-disruption ITER plasma using the newly developed non-linear code \NORSE, and \changed{describe a feedback mechanism by which} a transition to electron slide-away can be induced at field strengths significantly lower than previously expected. \changed{If the electric field is actively imposed using the control system, the entire electron population is quickly converted to runaways in the scenario considered. We find the time until the feedback mechanism sets in} to be highly dependent on the details of the mechanisms removing heat from the thermal electron population.
\end{abstract}

\section{Introduction}
Runaway electrons pose a severe threat to the safety and reliability of ITER and other high-plasma-current fusion devices \cite{hollmann15iter}. The larger the runaway-electron population, the larger the threat to the integrity of the device. However, if the electron momentum-space distribution function becomes highly non-Maxwellian due to the presence of a high-energy tail of runaways, existing numerical tools employing linearized collision operators are no longer valid. The same is true if the electric field (even momentarily) becomes comparable to the Dreicer field \cite{Dreicer1959}. 

We recently presented \NORSE \cite{Stahl2016b} -- an efficient solver of the kinetic equation in a homogeneous plasma -- which includes the full relativistic non-linear collision operator of Braams \&
Karney \cite{BraamsKarney1987,BraamsKarney1989}. \texttt{NORSE} -- which will be discussed in Section \ref{sec:NORSE} -- is able to model Dreicer and hot-tail runaway generation in the presence of electric fields of arbitrary strength and synchrotron-radiation reaction: one of the most important \changed{energy-}loss channels for runaways \cite{Stahl2015}. 

Since \NORSE is able to treat \changed{highly} distorted distributions, a range of new questions may be addressed. One issue of particular interest \changed{is:} will non-linear phenomena accelerate or dampen the growth of runaways? Naturally, this is of great importance in view of ITER and other large tokamaks, as it potentially impacts the requirements on the disruption mitigation system (the design of which is currently being finalized) \cite{hollmann15iter,Boozer2015}. In addition, the electric field is expected to reach values as high as 80-100 V/m during the current quench in ITER \cite{Smith2006}, and runaway generation is likely to be strong enough for 60\% or more of the plasma current to be converted to runaway current. In Section \ref{sec:ITER_disruption}, we use \NORSE to model the evolution of the electron population in a typical ITER post-disruption scenario. 

If the electric field is strong enough, the net \changed{parallel} force experienced by electrons due to the electric field and collisions becomes positive in the entire momentum space, leading to a phenomenon known as electron \emph{slide-away}. \changed{This is expected to happen when $E>0.215\ED\equiv \ESA$, where $\ED=ne^3\lnL/4\pi\epso^2 T$ is the so-called Dreicer field \cite{Dreicer1959}; $n$, $T$ and $-e$ are the
electron number density, temperature and charge; $\lnL$ is the Coulomb
logarithm; and $\epso$ is the vacuum permittivity.} The associated surge in runaway current can have a large impact on the potential for material damage, as well as the subsequent evolution of the parallel electric field. The slide-away process cannot be consistently modelled using linear tools such as \CODE \cite{CODEPaper2014,Stahl2016} or \LUKE \cite{Decker2004}, which assume a Maxwellian  background plasma and therefore require $E\ll\ESA$, as well as that the runaway fraction is small $n\sub{r}/n\ll 1$. 

A strong electric field represents a source of energy that quickly heats up the electron distribution. This heating can induce a transition to the slide-away regime -- even under a fixed applied electric field which is initially below the threshold $E<\ESA$ -- since the collisional friction is lower in a hotter distribution. As a consequence, the Dreicer field is also lower, making the effective normalized field $\EEDeff$ higher for a given field strength. If the temperature increase is large enough, the slide-away regime is reached, which happens at a field of $\EEDeff=0.215$ in the case of a constant applied electric field, i.e.~it coincides with the standard slide-away field at the effective temperature $T\sub{eff}$ \cite{Stahl2016b}. 

In practice, many processes act to remove heat from the plasma. In a cold post-disruption plasma, line radiation and bremsstrahlung from \changed{interactions with} partially ionized impurities are important loss channels, as is radial heat transport. Including a heat sink in numerical simulation of such scenarios is therefore desirable, and the sink effectively acts to delay or prevent the transition to slide-away. In this paper, we demonstrate that the evolution of the runaway electron population -- including the time to reach slide-away -- is highly sensitive to the properties of the applied heat sink, making a detailed investigation of the various loss channels an area of interest for future work.

\section{NORSE}
\label{sec:NORSE}
We will use the newly developed fully relativistic non-linear tool \NORSE \cite{Stahl2016b} to study the dynamics of the electron population. \NORSE, which is valid in spatially uniform plasmas, solves the kinetic equation
\begin{equation}
	\pder{f}{t} 
	-\frac{e\mathbf{E}}{\me c} 
		\cdot \pder{f}{\mathbf{p}} 
	+\pder{}{\mathbf{p}}\cdot
		\left(
			\mathbf{F}\sub{s}f
		\right)
	=C\sub{ee}\{f\} + C\sub{ei}\{f\} + S,
\end{equation}
where $f$ is the electron distribution function, $t$ is the time, $\me$ and $\mathbf{p}$ are the electron rest mass and momentum, $\mathbf{E}$ is the electric field, $c$ is the speed of light, $\mathbf{F}\sub{s}$ is the synchrotron-radiation-reaction force, $C\sub{ee}$ is the relativistic non-linear electron-electron collision operator, $C\sub{ei}$ is the electron-ion collision operator, and $S$ represents heat and particle sources or sinks. For a detailed description of the various terms and operators, see \Ref{Stahl2016b}. \changed{For the remainder of this paper, we define the electric field such that electrons are accelerated in the positive $p_{\|}$ direction.}

In \NORSE, the particle momentum $\mathbf{p}$ is represented in terms of the magnitude of the normalized momentum $p=\gamma v/c$ (where $v$ is the velocity of the particle and $\gamma$ is the relativistic mass factor) and the cosine of the pitch angle $\xi=p_{\|}/p$. The kinetic equation is discretized using finite differences in both $p$ and $\xi$. A linearly implicit time-advancement scheme is used, where the five relativistic \changed{Braams-Karney} potentials \cite{BraamsKarney1987} -- analogous to the Rosenbluth potentials in the non-relativistic case -- are calculated explicitly from the known distribution. These are then used to construct the electron-electron collision operator $C\sub{ee}$, and the remainder of the kinetic equation is solved implicitly.

For the results presented in this paper, a non-uniform finite-difference grid was used to improve the computational efficiency. In the pitch-angle coordinate, the grid points were chosen with a dense spacing close to $\xi=\pm 1$ -- in particular at $\xi=1$ where the runaway tail forms -- but with a sparser grid at intermediate $\xi$. In the $p$-direction, a grid mapping with a $\tanh$ step in spacing was used in order to produce a grid with dense spacing at low momenta (to accurately resolve the bulk dynamics), and larger spacing in the high-energy tail, where the scale length of variations in $f$ is larger. 

The runaway region in \NORSE was determined by studying particle trajectories in phase space, neglecting momentum-space diffusion but including self-consistent collisional friction and synchrotron-radiation reaction \cite{Stahl2016b}. The trajectory that terminates at $\xi=1$ and $p=p_c$ marks the lower boundary of the runaway region, since particles that follow it neither end up in the bulk nor reach arbitrarily high energies. The critical momentum $p_c$ is the momentum at which the balance of forces in the \changed{parallel direction} ($\xi=1$) becomes positive (i.e. the lowest momentum at which the accelerating force of the electric field overcomes the collisional and synchrotron-radiation-reaction drag). If the balance of forces is positive for all $p$ at $\xi=1$, however, all electrons experience a net acceleration, and the population is in the slide-away regime.

\subsection{Heat sink}
Including a heat sink \changed{(HS)} in the numerical simulations is of great importance for accurate modelling of the distribution evolution during a disruption. The heat sink used in \NORSE to remove heat from the thermal population has the form $S = \partial/\mathbf{\partial p} \cdot \big( k\sub{h} \mathbf{S}\sub{h} f \big)$, where $\mathbf{S}\sub{h}(p)$ is an isotropic function of momentum \changed{(i.e. $\mathbf{S}\sub{h}\, \|\, \mathbf{p}$)} and \changed{$k\sub{h}(t)$} is the magnitude of the source. The terms in the kinetic equation that affect the total energy content are the electric-field and synchrotron-radiation-reaction terms\changed{, however; when considering a subset $\boldsymbol{\Omega}$ of momentum space, collisions can also transfer energy in or out of $\boldsymbol{\Omega}$, and a corresponding term must be included. The} total energy change $\d W/\d t$ \changed{in $\boldsymbol{\Omega}$ can thus} be written as
\begin{equation}
\frac{\d W}{\d t} 
= \me c^2 \int_{\boldsymbol{\Omega}}\dthreep\, (\gamma-1)\!
	\left(
		-\frac{e\mathbf{E}}{\me c} 
		\cdot \pder{f}{\mathbf{p}} 
		+\pder{}{\mathbf{p}}\cdot
			\left(
				\mathbf{F}\sub{s}f
			\right)
		\changed{- C\{f\}}
		+k\sub{h}\pder{}{\mathbf{p}}\cdot(\mathbf{S}\sub{h}f)
	\right).
\label{eq:dWdt}
\end{equation}
\changed{The magnitude $k\sub{h}$ of the sink in each time step can be determined by requiring $\d W/\d t=0$. In this work, we take $\boldsymbol{\Omega}$ to represent the thermal bulk of the distribution, which we define as all particles with $v\!<\! 4 v\sub{th,0}$, with $v\sub{th,0}=\sqrt{2T_0/\me}$ the thermal speed at the initial temperature $T_0$.}

In this study, \changed{the $p$-component of $\mathbf{S}\sub{h}$} was chosen to \changed{have the shape of} a Maxwellian at the desired temperature $T$\changed{. In} practice, the momentum dependence of the sink will be more complicated and subject to the details of the particular physical processes at work. It will also likely have a limited energy-removal rate, dictated by for instance spatial gradients or impurity content, which could limit its efficiency in maintaining a given temperature. A detailed investigation of the characteristics of the \changed{sink} is left for future work; the \changed{aim} of this paper is to highlight the sensitivity of the runaway-electron evolution to the particulars of the sink\changed{, and for that purpose we will impose a limit on the energy-removal rate, as will be discussed in the next section}.

\section{Runaway generation in an ITER disruption}
\label{sec:ITER_disruption}

\subsection{Post-disruption scenario}

\begin{figure}
\begin{center}
\includegraphics[width=0.5\textwidth, trim={0.1cm 0cm 0.2cm 0.35cm},clip]{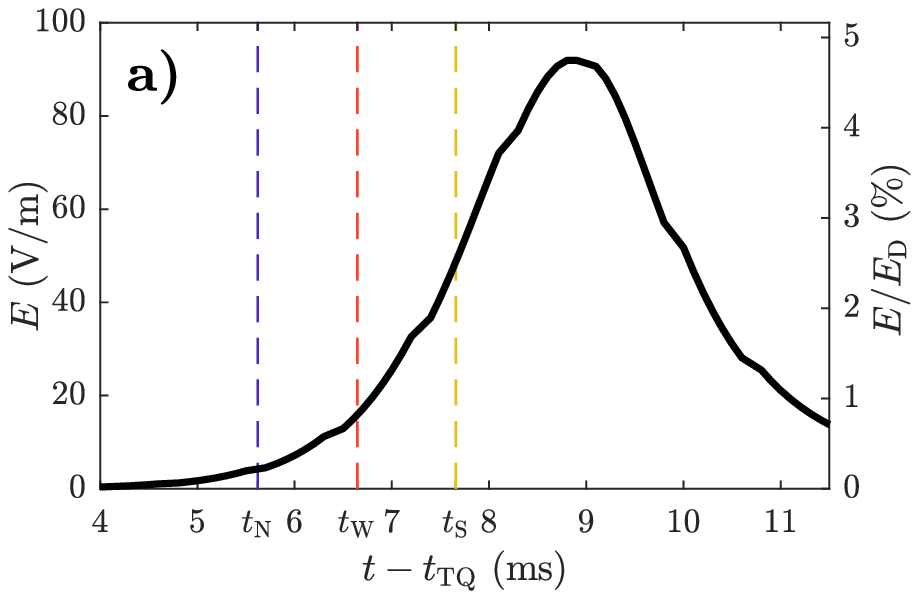}\hfill
\includegraphics[width=0.48\textwidth, trim={0.04cm 0cm 0.7cm 0.47cm},clip]{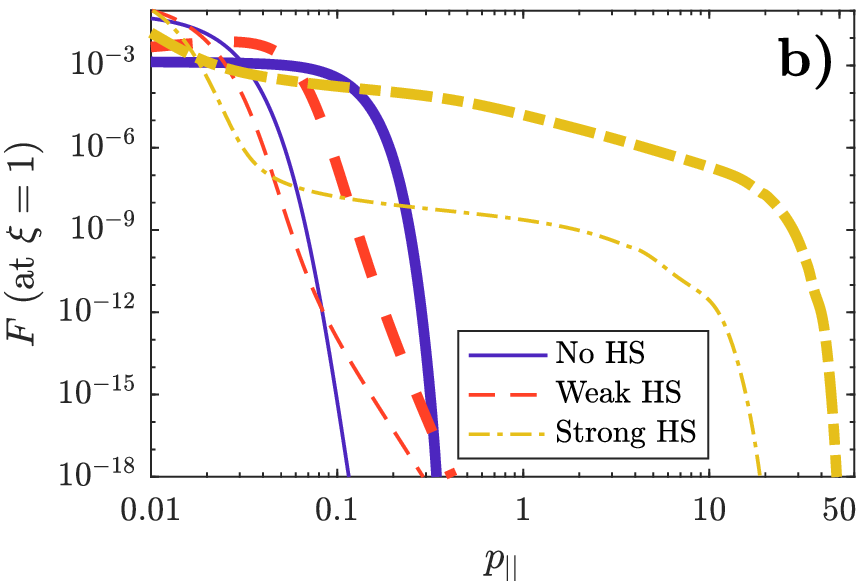}
\caption{\changed{a) Electric field in V/m (left vertical axis) and normalized to the Dreicer field $E\sub{D}$ at the temperature $T_0$ and density $n_0$ (right vertical axis), as a function of time after the thermal quench. b) Tail of the parallel electron distribution. Thin lines show $f$ at $t\sub{N}$ (no HS), $t\sub{W}$ (weak HS) and $t\sub{S}$ (strong HS), and thick lines show $f$ immediately before the transition to slide-away.}}
\label{fig:EFieldAndDist}
\end{center}
\end{figure}

In this section, we use \NORSE to model the \changed{evolution of the electron distribution} during a typical ITER disruption. The electric field evolution \changed{(which is shown in \Fig{fig:EFieldAndDist}a)} and other parameters are taken from the ITER inductive scenario no.~2 in \Ref{Smith2006}, \changed{but the temperature evolution has been simplified to facilitate the numerical calculation. We assume the electron population to be completely thermalized and use the final temperature $T=T_0=10$ eV throughout our simulation, together with the density $n=n_0=7.1\power{19}$ m$^{-3}$. This is likely to underestimate the runaway-electron generation in the early phase, in which the temperature is still dropping, however the chosen set-up is sufficient for our purposes. We use the magnetic field on axis ($B=5.3\,$T) and $\zeff=1$. The initial current density in the scenario is $j_0=0.62\,$MA/m$^2$, however our calculations start from a Maxwellian distribution and make no attempt to maintain the experimental current evolution explicitly.}

To highlight the importance of the temperature evolution of the bulk, in Section \ref{sec:results} we will consider three scenarios: \emph{no heat sink} (subscript N), \emph{weak heat sink} (W) and \emph{strong heat sink} (S). In the no-heat-sink scenario, all the energy supplied by the electric field will remain in the simulation, \changed{leading to rapid} bulk heating; with the strong heat sink, a bulk temperature of $T_0=10\,$eV will be \changed{enforced in accordance with Eq.~\ref{eq:dWdt},} i.e. any excess heat in the bulk region will be removed using a heat sink. In the intermediate case of a weak \HS, \changed{the energy-removal rate of the heat sink will be restricted to 0.5 MW/m$^3$. This particlar value has been chosen at will, but is meant to represent some inherent limitation in the physical processes responsible for the energy loss}. In both the weak and strong cases, the heat sink will affect only the thermal population, allowing the supra-thermal tail to gain energy from the electric field. Physically, this corresponds to processes not included in the simulation (such as radial transport or radiative losses) which primarily affect the thermal population.

\NORSE simulations of the evolution of the electron distribution function in the presence of the electric field in \Fig{fig:EFieldAndDist}a were performed for the three different scenarios. The simulations were aborted \changed{when} the runaway population reached $n\sub{r}/n=1$; i.e.~a transition to the slide-away regime was observed\changed{. The simulation results can however only be considered characteristic of a natural ITER disruption for current densities comparable to, or somewhat larger than, the initial value $j_0$,} since after that point the strong \changed{response  of the inductive electric field to the increased local current} would invalidate the $E$-field evolution used. \changed{We will therefore mark the time where the current density reaches $j/j_0>5$ in all plots, and denote it with $t\sub{N}$, $t\sub{W}$ and $t\sub{S}$, respectively, for the no-sink, weak-sink and strong-sink scenarios. The distribution evolution at later times can only be considered accessible in scenarios where the loop voltage is actively sustained using the control system. Nevertheless, this regime will turn out to be of interest, since a non-linear feedback mechanism leading to a rapid transition to slide-away is observed.}

\subsection{Evolution of the runaway-electron population}
\label{sec:results}
\changed{In \Fig{fig:EFieldAndDist}b, the tails of the distributions in the parallel direction are shown at $t\sub{N}$, $t\sub{W}$ and $t\sub{S}$ (thin lines), as well as at the final times (thick lines) in each scenario (just before the transition to slide-away is reached).} In the figure, the distribution is normalized such that $F=f/f_M(t\!=\!0,p\!=\!0)$, where \changed{$f\sub{M}$ is a Maxwell-J\"uttner distribution}, so that $F$ initially takes the value unity at $p=0$. The maximum achieved particle energies are highly dependent on whether a \HS was applied or not; in the no-heat-sink \changed{and weak-heat-sink scenarios}, the particles did not have time to reach relativistic energies, whereas in the \changed{strong \HS case, $p\approx 19$} and $p\approx 44$ (corresponding to energies of roughly \changed{$9$} and $22\,$MeV)\changed{,} were obtained \changed{at $t\sub{S}$ and just before reaching slide-away, respectively}. The reason for this is that, as we shall see, the \changed{current density growth and subsequent transition to slide-away} in the latter case occur at much later times, and the runaways have time to gain more energy. 

\begin{figure}
\begin{center}
\includegraphics[width=0.49\textwidth, trim={0.0cm 0cm 0.8cm 0.25cm},clip] {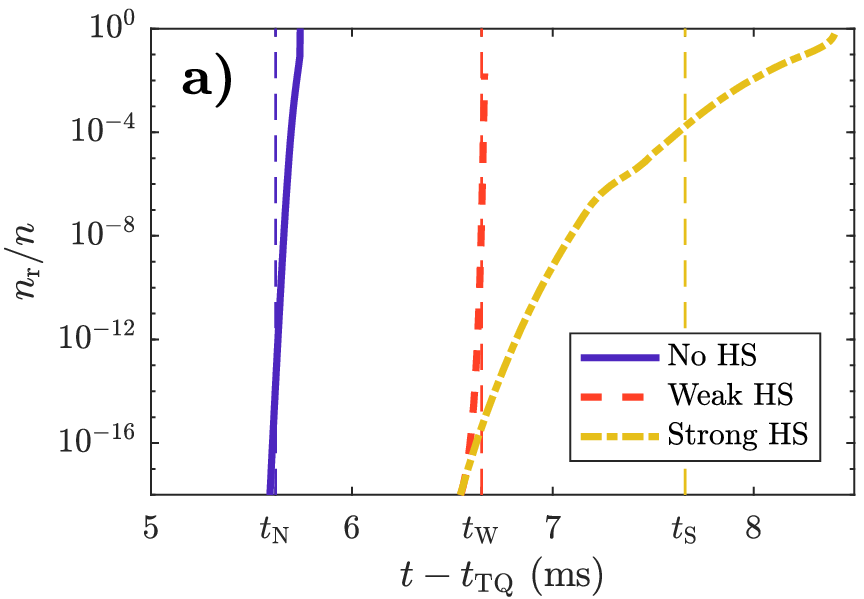}\hfill
\includegraphics[width=0.49\textwidth, trim={0cm 0cm 0.9cm 0.45cm},clip]{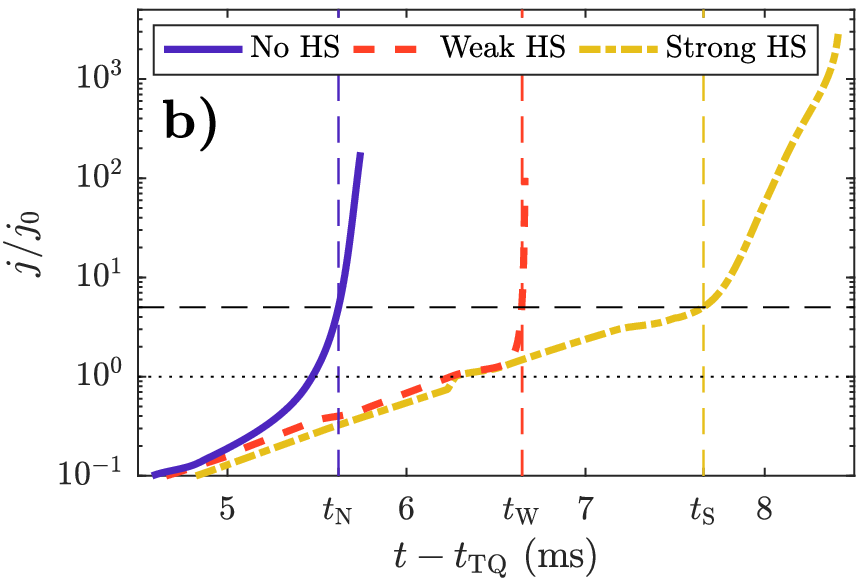}
\caption{\changed{a) Runaway fraction and b) current density normalized to its initial value $j_0$, as a function of time after the thermal quench in the different heat-sink scenarios. The times $t_{N}$, $t_{W}$ and $t_{S}$ (vertical thin dashed lines), mark the time where the current density reaches $j/j_0>5$ for the no-heat-sink, weak-heat-sink and strong-heat-sink scenarios, respectively.}}
\label{fig:jAndNr}
\end{center}
\end{figure}

Figure \ref{fig:jAndNr}a shows the evolution of the runaway fraction during the course of the simulation. In the no-heat-sink case, the runaway fraction increases sharply, but as shown in \changed{\Fig{fig:EFieldAndDist}b}, the runaways are all at low energy. The transition to slide-away happens already at \changed{$t$}$=5.7\,$ms; early on in the electric-field evolution (cf.~\Fig{fig:EFieldAndDist}). In the two scenarios employing a heat sink, the \changed{growth in runaway fraction occurs later, but in the weak-heat-sink case the growth rate is comparable to the case when a sink is absent once the process is initiated. In this case, the transition to slide-away happens at $t=6.7\,$ms.} With the strong \HS, the runaway population grows steadily, eventually dominating the entire distribution at \changed{$t=8.4\,$ms}, however in this case the transition is gradual, rather than rapid. In all three scenarios, including the \changed{one} with an ideal strong heat sink, the slide-away regime is thus reached even before the \changed{electric field} (calculated assuming a linear treatment) has reached its peak.

\changed{Figure \ref{fig:jAndNr}b shows the evolution of the current density. It indicates that the rapid increase in the runaway fraction is correlated with a similar increase in the current density, although in the no-sink case, the growth rate is somewhat smaller. Again, the growth in the strong-heat-sink case is gradual, rather than explosive.} Note that in the \changed{no and} weak heat-sink cases, the runaway fraction is still \changed{negligibly small} at the start of the rapid transition to slide-away. The transition is thus not a non-linear phenomenon triggered by the size of the runaway population; it starts in a regime where linearized tools are normally expected to be valid\changed{, and before the current density becomes significantly larger than its initial value}.

\begin{figure}
\begin{center}
\includegraphics[width=0.491\textwidth, trim={0.04cm 0cm 0.7cm 0.27cm},clip]{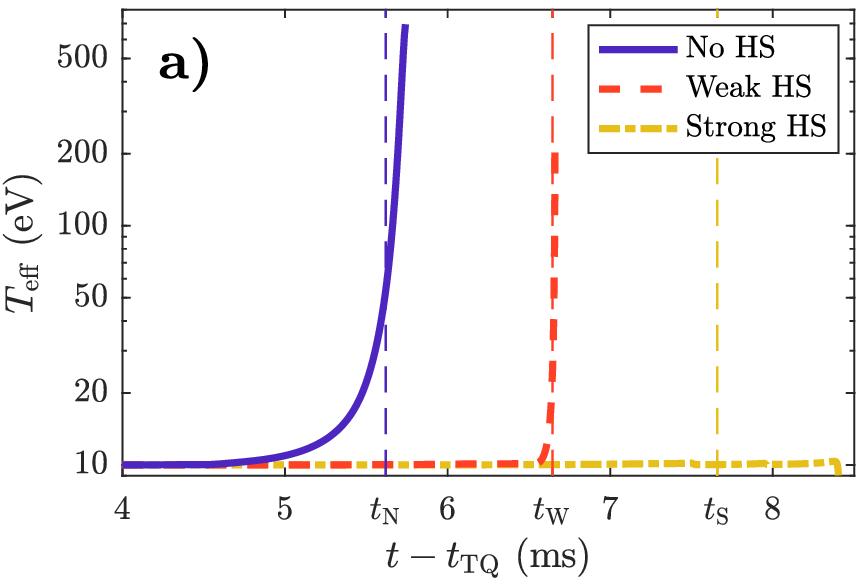}\hfill
\includegraphics[width=0.49\textwidth, trim={0.0cm 0cm 0.8cm 0.25cm},clip]{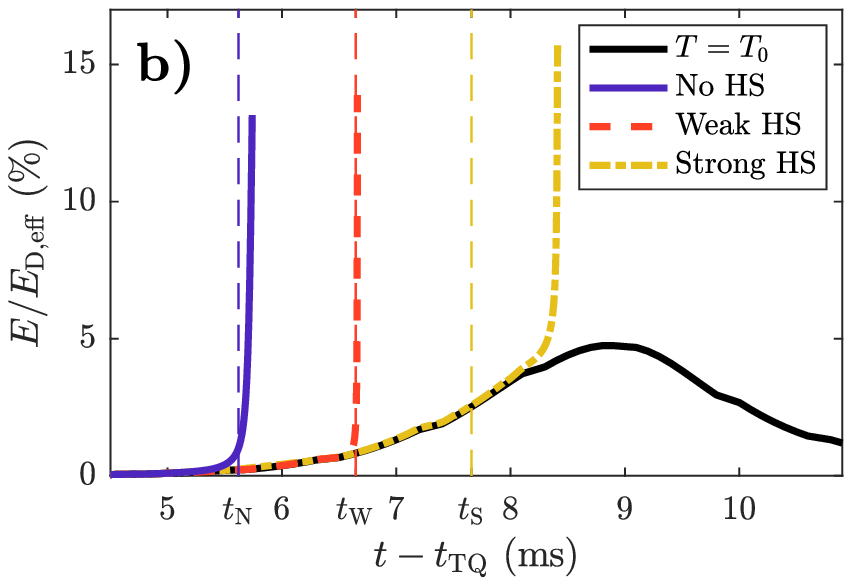}
\caption{a) Effective temperature of the bulk population and b) corresponding effective normalized electric-field strength \changed{(also taking changes to the bulk density into account)} as functions of time, in the different scenarios. The black solid line in panel b) corresponds to the normalized field with \changed{$T\!=\!T_0$ and $n\!=\!n_0$} (cf. \Fig{fig:EFieldAndDist}).}
\label{fig:EffectiveTAndEED}
\end{center}
\end{figure}

The explanation can be found by examining the thermal population. By comparing the energy moment of the bulk of the distribution \changed{($W\sub{\mathbf{\Omega}}$) with that of a relativistic Maxwellian ($W\sub{M}(T)$),
\begin{equation}
W\sub{\mathbf{\Omega}} 
= \me c^2 \int_{\mathbf{\Omega}} \!d^3p\, (\gamma-1) f
= W\sub{M} 
= \frac{\me c^2 n}{\Theta\sub{eff} K_2(1/\Theta\sub{eff})} \int_0^{p\sub{max,\mathbf{\Omega}}} \!\!\!\!\d p\, p^2 (\gamma-1) \exp\left( -\frac{\gamma}{\Theta\sub{eff}}\right),
\end{equation}
an effective temperature $T\sub{eff}$ for a given distribution $f$ can be determined by solving for $\Theta\sub{eff}=T\sub{eff}/\me c^2$. In the above equation, $p\sub{max,\mathbf{\Omega}}$ is the upper boundary in $p$ of the bulk region in momentum space, and $K_2(x)$ is the modified Bessel function of the second kind (and order two).
The effective temperature $T\sub{eff}$} is plotted in \Fig{fig:EffectiveTAndEED}a as a function of time for all three scenarios. In the no-\HS case, \changed{it} increases by \changed{roughly} two orders of magnitude during the simulation, i.e.~when energy is not actively removed from the system, the Ohmic heating is sufficient to heat the plasma to a temperature of \changed{about 700 eV} before the onset of slide-away\changed{, or 55 eV if the electric field is not artificially sustained}. A similar (albeit \changed{weaker}) tendency is seen in the weak-heat-sink case, where the temperature increases to \changed{$T\sub{eff}\approx 210\,$eV} in the phase leading up to the \changed{slide-away} transition\changed{, or 25 eV in a non-driven case}. \changed{This heating is a consequence of the imposed limited maximum energy-removal rate of the heat sink in the weak case, since the temperature is efficiently kept constant in the beginning of the simulation, where $(\d W/\d t)\sub{HS}<0.5\,$MW/m$^3$.} The strong \HS manages to keep $T\sub{eff}-T_0$ to within a few tenths of an eV during the entire simulation, corresponding to a source with unlimited (or at least higher than required) maximal energy-removal rate.

The significance of the observed bulk heating is its influence on the Dreicer field $\ED$, which (apart from the weak dependence on $\lnL$) is inversely proportional to $T$. For a given electric-field strength, the normalized field $\EED$ thus increases as the bulk heats up. The effective normalized electric field is shown in \Fig{fig:EffectiveTAndEED}b, indicating that the \changed{rapid growth in the runaway fraction in \Fig{fig:jAndNr}a} is correlated with a sudden increase in the normalized electric field in the no-heat-sink and weak-heat-sink cases. In the strong-heat-sink case, \changed{the increase in normalized field is not caused by the temperature, which is kept constant during the entire simulation, but by the decrease in the bulk density as the runaway population becomes substantial. As can be seen from the yellow dash-dotted line in the figure, this has a similar effect as a temperature increase, since $\ED\!\sim\! n\sub{bulk}$. The effective $\EED$ starts to deviate from the baseline value (black solid line) already when $n\sub{bulk}/n_0\approx 0.97$. The feedback process is thus initiated when the runaway fraction is just 3\%; a regime where linear tools are expected to be valid.} 

\changed{The two effects \changed{of increasing $T\sub{eff}$ and decreasing $n\sub{bulk}$} lead to a positive feedback mechanism which is responsible for the rapid growth in runaway fraction and current density seen in the no-heat-sink and weak-heat-sink cases. Once the bulk temperature has increased enough (or a high enough runaway tail is produced, as seen in the strong-heat-sink case), the normalized electric field becomes strong enough to cause a depletion of the bulk \changed{through primary runaway generation}. The reduced bulk density in turn leads to a more efficient heating of the remaining bulk particles. Both of these effects contribute to a reduction in the Drecier field $\ED$, and a corresponding reduced collisional friction on the bulk electrons, which makes runaway acceleration easier. This further increases the rate of bulk depletion, and so on.} Eventually the friction becomes low enough that the \changed{parallel} balance of forces becomes positive everywhere, marking the transition to the slide-away regime. At this point, the bulk of the distribution can no longer be well described by a Maxwellian and the positive feedback mechanism makes the transition possible even though $\EEDeff\!<\!0.215$ (in this case at around $\EEDeff\!\approx\!0.15$).

\begin{figure}
\begin{center}
\includegraphics[width=0.51\textwidth, trim={0.18cm 0cm 0.98cm 0.42cm},clip]{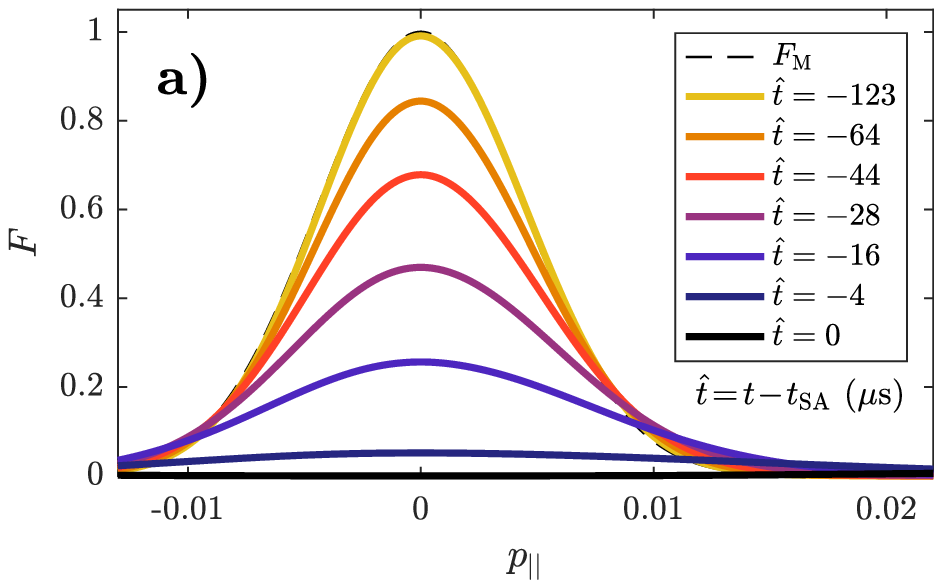}\hfill
\includegraphics[width=0.45\textwidth, trim={0cm 0cm 0.88cm 0.48cm},clip]{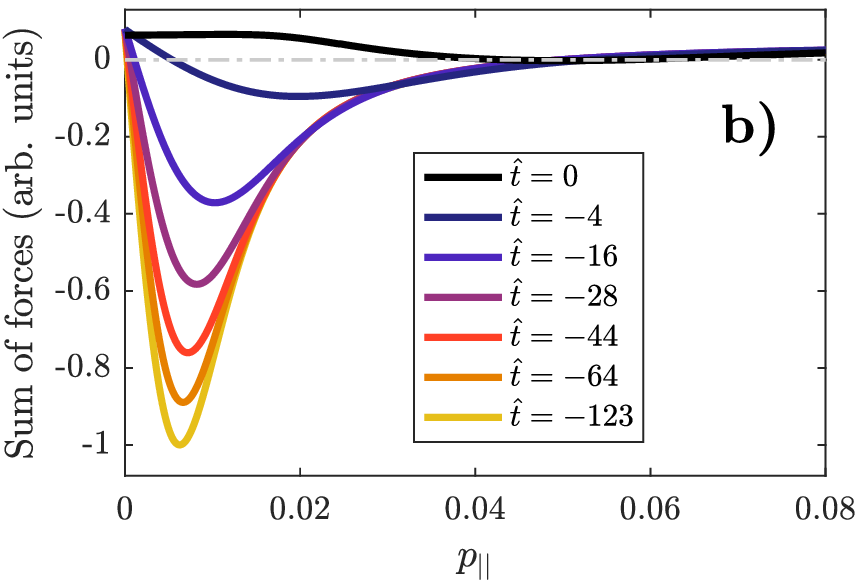}
\caption{a) Evolution of the bulk of the electron distribution and b) balance of forces, in the direction parallel to the electric field, just before the transition to slide-away in the weak-heat-sink scenario. $\hat{t}$ is the time relative to the transition to slide-away in \changed{$\mu$s}.}
\label{fig:BulkLoss}
\end{center}
\end{figure}

The bulk depletion and associated change in the \changed{parallel} force balance is shown in \Fig{fig:BulkLoss}a and \Fig{fig:BulkLoss}b, respectively, in the phase leading up to the transition to slide-away in the weak-heat-sink case. Initially, the force balance is positive in the tail -- i.e. particles there experience a net acceleration -- while it is negative in the bulk, meaning particles are slowed down by collisions. As the feedback process starts, the minimum in the sum of forces becomes gradually less pronounced, in tandem with the depletion of the bulk population, up until the point where the sum of forces becomes positive everywhere, and the slide-away regime is reached. 
This highly non-linear process cannot be \changed{accurately} captured by a linear model.

\section{Discussion and conclusions}
In this paper we have examined the \changed{evolution of the electron distribution} and runaway generation in an ITER\changed{-like} post-disruption scenario where the electric fields reach values as high as 90 V/m. With the help of the newly developed tool \NORSE, which is a relativistic non-linear solver for the electron momentum-space distribution function, we have shown that the slide-away regime, i.e.~a net \changed{parallel} acceleration of electrons in all of momentum space, is reached in this scenario\changed{, provided the electric field evolution used is artificially enforced by the control system}. \changed{In the stage leading up to the transition, a positive feedback mechanism sets in by means of which the bulk quickly gets depleted by primary runaway generation which reduces the friction on the thermal population, leading to further bulk depletion, until the point where the slide-away is reached. This process can be initiated at significantly weaker fields than the slide-away field $\ESA$ expected from linear theory.}

The time to a transition to slide-away is highly dependent on the ability of \changed{loss processes} to remove heat from the thermal electron population, but even with an ideal sink (the strong-heat-sink case in Section \ref{sec:ITER_disruption}), complete runaway generation was seen 8.4 ms after the thermal quench. These results were obtained without taking avalanche \changed{or hot-tail} runaway generation into account, which would only lead to more prominent runaway growth. 

\changed{Also in the case of a disruption where the electric field is not artificially sustained, strong bulk heating leads to a rapid growth in the runaway fraction because of the increase in the normalized electric field $\EED$, but the current density becomes large enough to significantly affect the electric field evolution (supressing the growth of $E$) before the slide-away regime is reached. This is observed in the absence of a heat sink, as well as with a heat sink with a limited maximum energy-removal rate (in this case 0.5 MW/m$^3$). If the efficiency of the heat sink is not limited, the runaway fraction grows more slowly and the runaways have time to reach significantly higher energies before the electric field becomes affected by the growing current density. The severity of disruptions in ITER could thus be greatly affected by the properties of the heat sinks present in the plasma.}

The feedback \changed{mechanism described} in this paper has \changed{important} consequences for the understanding of runaway-electron dynamics. With the entire electron population experiencing a net accelerating force at much weaker electric fields than previously expected, very large runaway-electron current generation is likely. This would impact the subsequent electric-field evolution, leading to a reduction in field strength and duration \changed{which could occur at realatively early times if the heat sink has a limited energy removal rate}. Therefore, it is difficult to determine the magnitude of the effect on the current evolution and post-quench dynamics without a self-consistent calculation of the electron distribution and the electric field. Nevertheless, this paper shows that \changed{feedback} effects play an important role in post-disruption runaway dynamics, and that the details of the heat-loss channels may have a big impact on what strength and duration of electric field can be tolerated \changed{before the positive feedback, and possible subsequent transition to slide-away, is induced}.

\vspace{-0.1cm}
\ack
This work was supported by the Swedish
Research Council (Dnr. 2014-5510), the
European Research Council (ERC-2014-CoG grant 647121), and
the Knut and Alice Wallenberg Foundation. M.L. was supported by the U.S. Department of Energy, Office of Science, Office of Fusion Energy Science, under Award Numbers DE-FG02-93ER54197 and DE-FC02-08ER54964.

\vspace{-0.1cm}
\section*{References}
\bibliography{Varenna16}

\end{document}